\documentclass[11pt,a4paper]{article}

\usepackage[margin=27mm]{geometry}
\usepackage{amsmath,amssymb}
\usepackage{booktabs}
\usepackage{graphicx}
\usepackage{authblk}
\usepackage{apacite}
\usepackage{url}
\usepackage{microtype}

\let\citeA\shortciteA

\title{The Alpha--Omega Problem in the Assessments of Atmospheric Power}
\author[1,2,3]{Anastassia M. Makarieva}
\author[1,2,3]{Andrei V. Nefiodov}
\affil[1]{Theoretical Physics Division, Petersburg Nuclear Physics Institute,
  Gatchina 188300, St. Petersburg, Russia}
\affil[2]{Institute for Advanced Study, Technical University of Munich,
  Lichtenbergstrasse 2 a, D-85748 Garching, Germany}
\affil[3]{Biotic Pump Greening Group Institute,
  S\~ao Jos\'e dos Campos, Brazil}
\date{\small Correspondence: \texttt{ammakarieva@gmail.com}}
\begin{document}
\maketitle

\begin{abstract}
Atmospheric power is commonly assessed with two Lorenz-cycle formulations treated
as equivalent: the $\alpha\omega$ formulation, yielding gas expansion power
$W_\omega$, and the $\mathbf{v}\cdot\nabla z$ formulation, yielding kinetic-energy
generation by horizontal pressure gradients, $W_K$. We show that, for a statistically stationary
global moist atmosphere, $W_\omega=W_K+W_P+W_{\rm nh}$, where $W_P$ is
the power of lifting precipitating water and $W_{\rm nh}$ is power associated with
departures from gas-phase hydrostatic balance. Since $W_P\simeq1$~W~m$^{-2}$
globally, equality would require $W_{\rm nh}\simeq-W_P<0$, so the associated vertical force imbalance would have to do negative work
globally. Using six-hourly ERA5, JRA-3Q, and MERRA-2 fields for 2025, we find that resolved $W_\omega$ varies by a factor of 1.7
(1.9--3.2~W~m$^{-2}$), versus 1.2 for $W_K$ (2.4--2.8~W~m$^{-2}$). Zonal-mean
$\alpha\omega$ profiles have area-weighted contributions near
$\pm50$~W~m$^{-2}$, making global $W_\omega$ a cancellation residual, whereas
$\mathbf{v}\cdot\nabla z$ profiles vary much less with latitude. The difference
$W_\omega-W_K$ ranges from $-0.61$ to $+0.87$~W~m$^{-2}$. Thus reanalysis-derived $W_\omega$ cannot
provide an independent constraint on atmospheric power. Two treatments of the omitted surface layer 
give cross-reanalysis mean $W_K$ of 2.8 and 3.1~W~m$^{-2}$, with
corrected values spanning 2.6--3.5~W~m$^{-2}$. Both exceed the observation-based
Lorenz available-potential-energy generation estimate of 1.2~W~m$^{-2}$.
Assuming $W_{\rm nh}\geq0$, adding $W_P\simeq1$~W~m$^{-2}$ gives total
atmospheric power of about 3.8--4.1~W~m$^{-2}$, motivating investigation of
additional power-generation mechanisms.

\end{abstract}

\section{Introduction}
\label{sec:introduction}
The rate at which the atmosphere generates mechanical power is a basic measure of
its circulation intensity and provides a quantitative constraint on the mechanisms
that maintain it. Published estimates of atmospheric power, however, differ
substantially in magnitude and do not always refer to the same physical quantity.

Most quantitative estimates have been made within the Lorenz energy cycle (LEC)
\cite{Lorenz1955,Oort1964,Li2007,BoerLambert2008,Marques2009,KimKim2013,
Pan2017,Ma2021,Oh2026}. In the Lorenz framework, differential diabatic heating
generates available potential energy (APE), which is subsequently converted into
kinetic energy by atmospheric motion. For a statistically stationary global
atmosphere, the total generation of APE must balance its net conversion to kinetic
energy. In the classical four-reservoir cycle, this
net conversion is
\begin{equation}
  W_{\rm LEC}\equiv C(P_M,K_M)+C(P_E,K_E).
  \label{eq:WLEC}
\end{equation}
Here $W_{\rm LEC}$ is the global mean APE-to-kinetic-energy conversion rate per
unit surface area; $P_M$ and $K_M$ are the zonal-mean APE and kinetic-energy
reservoirs, respectively; $P_E$ and $K_E$ are the corresponding eddy APE and
kinetic-energy reservoirs; and $C(X,Y)$ denotes the conversion rate from
reservoir $X$ to reservoir $Y$. The corresponding APE generation is
$G_{\rm LEC} \equiv G(P_M)+G(P_E)$, where $G(P_M)$ and $G(P_E)$ denote diabatic generation of
zonal-mean and eddy APE, respectively. In statistical steady state the classical
LEC requires $G_{\rm LEC}=W_{\rm LEC}$.

The conversion side of the LEC, $W_{\rm LEC}$, has commonly been evaluated using
two formulations. The
first is the pressure-velocity form,
\begin{equation}
  W_\omega
  \equiv -\frac{1}{\mathcal{S}}\int_V \omega\,\mathrm{d}V,
  \qquad \omega\equiv\frac{Dp}{Dt},
  \label{eq:Womega}
\end{equation}
where $\mathcal{S}$ is Earth's total surface area, $V$ is atmospheric volume,
$p$ is air pressure, and $D/Dt$ is the material derivative. For the statistically
stationary global means considered here, with negligible
pressure-work flux through the control boundary, this quantity represents
gaseous expansion power. Under hydrostatic balance it can be written on
pressure levels as
\begin{equation}
  W_\omega
  =-\frac{1}{\mathcal{S}g}\int_S\int_{p_t}^{p_s}
  \alpha\omega\,\mathrm{d}p\,\mathrm{d}S,
  \label{eq:Womega_pressure}
\end{equation}
where $\alpha\equiv 1/\rho$ is the specific volume of gaseous air, $\rho$ is its
density, $g$ is gravitational acceleration, $S$ denotes Earth's surface, and
$p_t$ and $p_s$ are the top and surface pressures.
Equation~\eqref{eq:Womega_pressure} is the alpha--omega form commonly used in
LEC calculations.

The second formulation gives the rate of kinetic-energy generation by horizontal
pressure gradients,
\begin{equation}
  W_K
  \equiv-\frac{1}{\mathcal{S}}\int_V
  \boldsymbol{u}\boldsymbol{\cdot}\boldsymbol{\nabla}p\,\mathrm{d}V,
  \label{eq:WK}
\end{equation}
where $\boldsymbol{u}$ is horizontal air velocity and $\boldsymbol{\nabla}p$ is the
pressure gradient. Transforming the volume
integral to pressure coordinates gives
\begin{equation}
  W_K
  =-\frac{1}{\mathcal{S}g}\int_S\int_{p_t}^{p_s}
  \boldsymbol{u}\boldsymbol{\cdot}\boldsymbol{\nabla}_p\Phi
  \,\mathrm{d}p\,\mathrm{d}S,
  \label{eq:WK_pressure_intro}
\end{equation}
where $\boldsymbol{\nabla}_p$ denotes differentiation with respect to horizontal
position at fixed pressure and $\Phi$ is geopotential.
Unlike Eq.~\eqref{eq:Womega_pressure}, this transformation does not require
hydrostatic balance.
Equation~\eqref{eq:WK_pressure_intro} is commonly called the
$\mathbf{v}\boldsymbol{\cdot}\nabla z$ formulation in the LEC literature,
where $\mathbf{v}$ in this conventional label denotes horizontal velocity;
here we denote horizontal velocity by $\boldsymbol{u}$.

The generation and conversion sides of the cycle can in principle be estimated independently.
APE generation can be diagnosed from atmospheric temperature and diabatic
heating rates, the latter being constrained by radiative, sensible, and latent
heat-flux fields. The conversion side is constrained by wind, pressure,
geopotential, and temperature fields. Because global fields of diabatic heating are
much less directly constrained
observationally, most global LEC estimates have used the conversion side of the
budget \cite{RomanskiRossow2013}.
Reanalysis-based estimates of $W_{\rm LEC}$ from these pressure-level
formulations are generally around 2--3~W~m$^{-2}$, with appreciable dependence
on dataset and methodology
\cite{Li2007,Marques2009,KimKim2013,Ma2021,Oh2026}.
Outside the LEC framework, MERRA-based estimates include kinetic-energy
generation by horizontal pressure gradients, $W_K$, of about
2.5~W~m$^{-2}$ \cite{HuangMcElroy2015}, and a moist-atmospheric heat-engine
work output of about 3.7~W~m$^{-2}$ \cite{Laliberte2015}. In contrast, the observation-based APE-generation components reported by
\citeA{RomanskiRossow2013} sum to only
$G_{\rm LEC}=1.23$~W~m$^{-2}$. Whether these
differences reflect distinct physical components of atmospheric power or
differences among the diagnostics is therefore a basic issue for interpreting
the atmospheric energy budget.

For a statistically stationary dry hydrostatic atmosphere obeying source-free
continuity, complete global integration with consistent boundaries gives
$W_\omega=W_K$ \cite{Lorenz1955,PeixotoOort1992}. The two expressions
have therefore generally been treated as equivalent formulations of the same
Lorenz conversion. Nevertheless, discrepancies between their empirical estimates,
of varying magnitude, have been noted and discussed in several contexts when
analyzing pressure-level reanalysis data.
\citeA{Li2007} identified reanalysis pressure vertical velocity $\omega$ as
having relatively low data quality and, to take advantage of the higher data
quality of horizontal velocity, transformed $C(P_M,K_M)$ and $C(P_E,K_E)$
into horizontal-wind form using the continuity equation. \citeA{Marques2009} found
the two formulations to be comparable only when
below-ground pressure-level data were excluded from the calculations. 
\citeA{KimKim2013} obtained broadly comparable global conversions using the
alpha--omega and horizontal-wind formulations, although differences remained
among individual conversion terms.

Thus, these discrepancies have generally been interpreted as arising from data
quality or implementation rather than from a failure of the underlying
equivalence. Using an ECMWF forecast model, \citeA{Steinheimer2008} characterized
the two formulations as hydrostatically equivalent, compared them in selected
cases, and found their global averages identical within error
margins, although without reporting this analysis in detail. They then employed
the alpha--omega formulation exclusively. As discussed below, however, such a
comparison is itself conditioned by the continuity equation used by the model
to determine $\omega$. In the most recent multi-reanalysis assessment,
\citeA{Oh2026} likewise described the two formulations as theoretically
equivalent when globally integrated, but noted that the alpha--omega
formulation is more sensitive to uncertainty because of inaccuracies in
reanalysis $\omega$ and therefore adopted the
$\mathbf{v}\boldsymbol{\cdot}\nabla z$ formulation.

More serious inconsistencies appear in the alpha--omega estimates themselves.
From the conversion components reported by \citeA{KimChoi2017}, the net
APE-to-kinetic-energy conversion is $-3.54$~W~m$^{-2}$ for NCEP2 but
$+3.00$~W~m$^{-2}$ for ERA-Interim.
Independently, \citeA{Makarieva2017} likewise obtained negative global
pressure-velocity power from NCAR/NCEP and positive from MERRA. 
Thus alpha--omega estimates can differ even in sign among reanalyses, whereas
the horizontal-wind formulation is substantially more robust.

These results expose two distinct questions. First, are the alpha--omega and
horizontal-wind formulations physically equivalent in a moist atmosphere?
Second, what relation should be expected between the two diagnostics when they
are evaluated from reanalysis fields whose mass-continuity equation, archived
fields, and subsequent mass corrections need not reproduce physical gaseous-air
continuity?

The first question is theoretical. The usual transformation between
Eqs.~\eqref{eq:Womega} and \eqref{eq:WK} uses source-free continuity. In
gaseous moist air, evaporation and condensation are sources and sinks of gas
mass, and the vertical pressure gradient need not equal the weight of gaseous
air alone. For a statistically stationary global control volume, we derive
\begin{equation}
  W_\omega=W_K+W_P+W_{\rm nh},
  \label{eq:master_intro}
\end{equation}
where $W_P$ is the gravitational power associated with lifting water that
subsequently leaves the gas phase and $W_{\rm nh}$ is the power associated with
departure from gas-phase hydrostatic balance. The dry hydrostatic equality is
recovered when both terms vanish.

The second question concerns the continuity equation actually associated with
the reanalysis fields. Model pressure velocity is tied to the mass variable and
source convention of the underlying dynamical system, which need not coincide
with physical gaseous-air continuity. Archived fields can acquire additional
continuity inconsistencies through assimilation, interpolation, and temporal
sampling, while postprocessing procedures may impose physical dry-air
conservation and reconstruct $\omega$ from the corrected mass fluxes. Agreement
or disagreement between the two reanalysis diagnostics therefore need not
constitute a direct measurement of the physical terms in
Eq.~\eqref{eq:master_intro}.

This distinction is especially important because the global alpha--omega
integral is a small residual of much larger positive and negative contributions.
\citeA{Steinheimer2008} showed strong cancellation between these contributions,
so spatially structured changes in $\omega$ can be small relative to the local
terms yet large relative to the global residual.

Finite resolution introduces a further distinction between archived and total
physical power. Pressure-level fields omit kinetic-energy generation by motions
below their spatial and temporal resolution. They also omit the real partial layer
between the lowest valid pressure
surface and the ground. This can be consequential because prominent
kinetic-energy generation occurs in the lower atmosphere
\cite{HuangMcElroy2015}, while our calculations below show a pronounced
maximum near the surface.

Section~\ref{sec:theory} derives the relation between $W_\omega$ and $W_K$ for
gaseous moist air, including nonhydrostatic vertical-force work.
Section~\ref{sec:unresolved} examines how the mass-continuity conventions of
reanalysis systems, continuity errors in archived fields, and subsequent mass
corrections affect the relation between the two resolved diagnostics.
Sections~\ref{sec:methods} and \ref{sec:results} evaluate both diagnostics from
six-hourly ERA5, JRA-3Q, and MERRA-2 pressure-level fields for 2025. The analysis
is designed to determine which differences between published power estimates
reflect atmospheric physics and which reflect the construction and conditioning
of the diagnostic.

\section{Atmospheric power with phase changes and nonhydrostatic motion}
\label{sec:theory}
The relation between Eqs.~\eqref{eq:Womega} and \eqref{eq:WK} follows from mass
continuity and vertical force balance. Let
$\boldsymbol{v}=\boldsymbol{u}+w\boldsymbol{e}_z$ be the gas-phase velocity,
where $w$ is vertical velocity, $\boldsymbol{e}_z$ is the upward unit vector, and
$z$ is height. Since
$\omega=\partial p/\partial t+\boldsymbol{v}\boldsymbol{\cdot}\boldsymbol{\nabla}p$,
\begin{equation}
  W_\omega=W-B_p,
  \qquad
  W\equiv-\frac{1}{\mathcal{S}}\int_V
  \boldsymbol{v}\boldsymbol{\cdot}\boldsymbol{\nabla}p\,\mathrm{d}V,
  \qquad
  B_p\equiv\frac{1}{\mathcal{S}}\int_V
  \frac{\partial p}{\partial t}\,\mathrm{d}V,
  \label{eq:womega_total_work}
\end{equation}
where $W$ is total pressure-gradient power and $B_p$ is pressure storage. Splitting
$W$ into horizontal and vertical parts gives
\begin{equation}
  W=W_K-\frac{1}{\mathcal{S}}\int_V
  w\frac{\partial p}{\partial z}\,\mathrm{d}V.
  \label{eq:w_height_split}
\end{equation}

Gaseous moist air obeys
\begin{equation}
  \frac{\partial\rho}{\partial t}
  +\boldsymbol{\nabla}\boldsymbol{\cdot}(\rho\boldsymbol{v})
  =\dot{\rho},
  \label{eq:cont_height_source}
\end{equation}
where $\dot{\rho}$ is the gas-phase mass source, positive for evaporation and
negative for condensation. Multiplication by $gz$ and integration over the
atmosphere gives
\begin{equation}
  \frac{1}{\mathcal{S}}\int_V\rho gw\,\mathrm{d}V=W_P+B_g,
  \label{eq:vertical_gravity_wp}
\end{equation}
with
\begin{equation}
  W_P\equiv-\frac{1}{\mathcal{S}}\int_Vgz\dot{\rho}\,\mathrm{d}V,
  \qquad
  B_g\equiv
  \frac{1}{\mathcal{S}}\frac{\mathrm{d}}{\mathrm{d}t}
  \int_V\rho gz\,\mathrm{d}V
  +\frac{1}{\mathcal{S}}\int_{\partial V}
  \rho gz\boldsymbol{v}\boldsymbol{\cdot}\boldsymbol{n}\,\mathrm{d}S,
  \label{eq:wp_bg}
\end{equation}
where $B_g$ contains gaseous gravitational-energy storage and boundary fluxes,
$\partial V$ is the control-volume boundary, and $\boldsymbol{n}$ is its outward
unit normal. 
For the global atmospheric volume, the boundary-flux term vanishes at the
surface because $z=0$ and at the top of the atmosphere as atmospheric mass
tends to zero.

For a statistically stationary global water cycle, let $P$ be the global-mean
precipitation mass flux and define the effective source--sink height
\begin{equation}
  H_P\equiv-\frac{1}{P\mathcal{S}}
  \int_V z\dot{\rho}\,\mathrm{d}V.
  \label{eq:hp_def}
\end{equation}
Then
\begin{equation}
  W_P=gPH_P.
  \label{eq:wp_gph}
\end{equation}
When evaporation occurs predominantly near the surface, $H_P$ can be interpreted
as an effective mean height of removal of water from the gas phase.

Define the departure from gas-phase hydrostatic balance by
\begin{equation}
  f_{\rm nh}\equiv-\frac{\partial p}{\partial z}-\rho g,
  \label{eq:fnh_def}
\end{equation}
and let
$W_{\rm nh}\equiv\mathcal{S}^{-1}\int_Vwf_{\rm nh}\,\mathrm{d}V$. Combining
Eqs.~\eqref{eq:womega_total_work}, \eqref{eq:w_height_split}, and
\eqref{eq:vertical_gravity_wp} gives
\begin{equation}
  W_\omega=W_K+W_P+W_{\rm nh}+B,
  \qquad
  B\equiv B_g-B_p.
  \label{eq:nonhydro_main}
\end{equation}
For a statistically stationary global control volume, $B=0$, giving
Eq.~\eqref{eq:master_intro}.

Here $W_{\rm nh}$ measures the global work done by departures from gas-phase
hydrostatic balance. Condensate loading may contribute to this departure, but
no decomposition of $W_{\rm nh}$ is required here. If $W_K>W_\omega$ while
$W_P>0$ and $B$ is negligible, then $W_{\rm nh}<-W_P$; this global condition
does not imply negative buoyancy everywhere.

Pressure can be used as a vertical coordinate without assuming hydrostatic
balance, provided that $p$ is monotonic with height. Define the
pressure-to-height Jacobian
\begin{equation}
  J_p\equiv
  -\left(\frac{\partial z}{\partial p}\right)_{x,y,t}>0.
  \label{eq:pressure_jacobian}
\end{equation}
Then $\mathrm{d}V=J_p\,\mathrm{d}p\,\mathrm{d}S$, and the volume-integral
expansion power becomes
\begin{equation}
  W_\omega
  =-\frac{1}{\mathcal{S}}\int_S\int_{p_t}^{p_s}
  J_p\omega\,\mathrm{d}p\,\mathrm{d}S.
  \label{eq:Womega_pressure_general}
\end{equation}
From Eq.~\eqref{eq:fnh_def},
\begin{equation}
  J_p=\frac{1}{\rho g+f_{\rm nh}}.
  \label{eq:pressure_jacobian_fnh}
\end{equation}
Only under gas-phase hydrostatic balance, $f_{\rm nh}=0$, does
$J_p=\alpha/g$, reducing Eq.~\eqref{eq:Womega_pressure_general} to the
conventional alpha--omega expression.

By contrast, the pressure-coordinate representation of $W_K$ does not
require hydrostatic balance. Implicit differentiation of
$p(x,y,z,t)=p$ at fixed pressure gives
\begin{equation}
  \boldsymbol{\nabla}_z p
  =-\frac{\partial p}{\partial z}\boldsymbol{\nabla}_p z,
  \label{eq:pressure_gradient_transform}
\end{equation}
where $\boldsymbol{\nabla}_z$ denotes the horizontal gradient at fixed
height. Since
$J_p=-(\partial p/\partial z)^{-1}$, the Jacobian cancels:
\begin{equation}
  J_p\boldsymbol{\nabla}_z p=\boldsymbol{\nabla}_p z.
  \label{eq:pressure_gradient_jacobian}
\end{equation}
Therefore
\begin{equation}
  W_K
  =-\frac{1}{\mathcal{S}}\int_S\int_{p_t}^{p_s}
  \boldsymbol{u}\boldsymbol{\cdot}\boldsymbol{\nabla}_p z\,
  \mathrm{d}p\,\mathrm{d}S
  =-\frac{1}{\mathcal{S}g}\int_S\int_{p_t}^{p_s}
  \boldsymbol{u}\boldsymbol{\cdot}\boldsymbol{\nabla}_p\Phi\,
  \mathrm{d}p\,\mathrm{d}S.
  \label{eq:WK_pressure_general}
\end{equation}

The same distinction appears in mass continuity. In general pressure
coordinates, define the mass Jacobian
\begin{equation}
  \mu_p\equiv\rho J_p.
\end{equation}
Equation~\eqref{eq:cont_height_source} then becomes
\begin{equation}
  \frac{\partial\mu_p}{\partial t}
  +\boldsymbol{\nabla}_p\boldsymbol{\cdot}(\mu_p\boldsymbol{u})
  +\frac{\partial(\mu_p\omega)}{\partial p}
  =J_p\dot{\rho}.
  \label{eq:cont_pressure_general}
\end{equation}
Under hydrostatic balance, $J_p=\alpha/g$ and $\mu_p=1/g$, so that
Eq.~\eqref{eq:cont_pressure_general} reduces to
\begin{equation}
  \boldsymbol{\nabla}_p\boldsymbol{\cdot}\boldsymbol{u}
  +\frac{\partial\omega}{\partial p}
  =\alpha\dot{\rho}.
  \label{eq:cont_pressure_source}
\end{equation}
Vertical integration with consistent boundary terms then reproduces
$W_\omega=W_K+W_P+B$, the hydrostatic limit of
Eq.~\eqref{eq:nonhydro_main}.

Thus departures from hydrostatic balance affect the two pressure-coordinate
diagnostics asymmetrically. They invalidate the conventional
$\alpha\omega$ representation of $W_\omega$ by changing the
pressure-to-volume Jacobian, whereas the corresponding Jacobian cancels
identically in the pressure-coordinate representation of $W_K$. The three
reanalyses analyzed below employ hydrostatic dynamical cores, so this
additional distinction does not affect the internal evaluation of their
pressure-level diagnostics; it becomes relevant when these diagnostics are
interpreted as measures of power in the actual atmosphere or applied to
nonhydrostatic model output.
Accordingly, these hydrostatic reanalyses cannot independently diagnose the
actual atmospheric $W_{\rm nh}$. In the comparisons below, $W_{\rm nh}$
enters through the physical identity for the atmosphere, not as a resolved
reanalysis diagnostic.

Hydrostaticity therefore removes the Jacobian ambiguity for the reanalyses,
but it does not remove a separate numerical issue: the alpha--omega
diagnostic is strongly cancellation dominated. This follows from the same
vertical pressure-work decomposition used above. Under hydrostatic balance,
\begin{equation}
  -\alpha\omega
  =
  gw-\alpha\left(
  \frac{\partial p}{\partial t}
  +\boldsymbol{u}\boldsymbol{\cdot}\boldsymbol{\nabla}_z p
  \right).
  \label{eq:alphaomega_local_split}
\end{equation}
Thus $-\alpha\omega$ contains the mass-specific vertical pressure-gradient
work $gw$, which changes sign between ascent and subsidence. Yet
Eq.~\eqref{eq:vertical_gravity_wp} constrains its global mass-weighted
integral to $W_P+B_g$. Large positive and negative vertical-work
contributions must therefore nearly cancel in the global mean. This
cancellation is distinct from the physical difference between $W_\omega$
and $W_K$ in Eq.~\eqref{eq:nonhydro_main}; it persists even in a dry
hydrostatic atmosphere.

The severity of this cancellation reflects the disparity between vertical
and horizontal pressure gradients. A characteristic vertical gradient is
\begin{equation}
  \left|\frac{\partial p}{\partial z}\right|
  \sim\frac{p_s}{H}
  \sim10^4\ \mathrm{Pa\ km^{-1}},
  \label{eq:vert_grad_scale}
\end{equation}
where $H$ is a pressure scale height, whereas a large-scale horizontal pressure
gradient is typically of order $1\ \mathrm{Pa\ km^{-1}}$. Even with
$|w|/|\boldsymbol{u}|\sim10^{-2}$, local vertical pressure-gradient power
density can exceed its horizontal counterpart by about two orders of magnitude.

\section{Unresolved motions and resolved mass continuity}
\label{sec:unresolved}
The preceding section showed that, in the physical atmosphere,
Eq.~\eqref{eq:nonhydro_main} relates $W_\omega-W_K$ to $W_P$,
$W_{\rm nh}$, and the storage and boundary term $B$. Since
Eq.~\eqref{eq:wp_bg} expresses $W_P$ as the geopotential moment of the
gas-phase mass source $\dot{\rho}$, the source or sink in gaseous mass
continuity enters the difference between the two power formulations directly.
In the hydrostatic limit, the physical difference between $W_\omega$ and
$W_K$ therefore depends on this mass source, apart from storage and boundary
terms.

The pressure velocity in a model or reanalysis, however, is tied to the
continuity equation of its resolved dynamical system rather than directly to
Eq.~\eqref{eq:cont_pressure_source}. To distinguish the two, let a superscript
${\rm r}$ denote resolved model or reanalysis fields, without implying a
Reynolds averaging operation, and write their hydrostatic pressure-coordinate
continuity equation schematically as
\begin{equation}
  \boldsymbol{\nabla}_p\boldsymbol{\cdot}\boldsymbol{u}^{\rm r}
  +\frac{\partial\omega^{\rm r}}{\partial p}
  =S^{\rm r},
  \label{eq:reanalysis_continuity}
\end{equation}
where $S^{\rm r}$ is the source term represented in the continuity equation
that determines the resolved mass transport. If the mass variable were gaseous
air and its physical source were represented, then
$S^{\rm r}=\alpha^{\rm r}\dot{\rho}^{\rm r}$. More generally this need not be
the case.

For a hydrostatic set of resolved fields whose pressure, mass measure, velocity,
and continuity equation are internally consistent, the same integration used
in Section~\ref{sec:theory} gives
\begin{equation}
  W_\omega^{\rm r}
  =W_K^{\rm r}+W_S^{\rm r}+B^{\rm r},
  \qquad
  W_S^{\rm r}
  \equiv
  -\frac{1}{\mathcal S g}\int_S\int_{p_t}^{p_s}
  \Phi^{\rm r}S^{\rm r}\,\mathrm{d}p\,\mathrm{d}S,
  \label{eq:reanalysis_power_identity}
\end{equation}
where $B^{\rm r}$ denotes the corresponding storage and boundary contributions.
Only when $S^{\rm r}=\alpha^{\rm r}\dot{\rho}^{\rm r}$ does
$W_S^{\rm r}$ have the meaning of a resolved gaseous $W_P$. In particular, a
source-free model continuity equation has $W_S^{\rm r}=0$ and, apart from
boundary and storage terms, restores the conventional hydrostatic equivalence
of the two resolved diagnostics.

The distinction is explicit in the ECMWF Integrated Forecasting System (IFS).
In the standard IFS formulation used before Cycle 46R1, the dynamical
continuity equation conserves total atmospheric mass rather than dry-air mass
\cite{Malardel2019}. Physical parameterizations can nevertheless change total atmospheric water
through surface exchange, precipitation, and subgrid transport.
Because these water-mass changes are not coupled to the standard dynamical
continuity equation, they imply equal and opposite changes of dry-air mass, so
that the total atmospheric mass source remains zero. Writing $S_w$ for such a
total-water source or sink, the implied constituent bookkeeping is
schematically
\begin{equation}
  S_d^{\rm art}=-S_w,
  \qquad
  S_t=S_w+S_d^{\rm art}=0.
  \label{eq:ifs_dry_air_source}
\end{equation}
Internal phase changes among the model water species cancel in the total-water
budget, so Eq.~\eqref{eq:ifs_dry_air_source} should not be interpreted as an
artificial dry-air source accompanying every condensation event. Its relevance
here is that the source-free total-mass continuity equation used by the
dynamical core does not contain the physical gas-phase source in
Eq.~\eqref{eq:cont_pressure_source}.

This distinction provides a natural interpretation of the equivalence reported
by \citeA{Steinheimer2008}. Their calculations used IFS Cycle 32R3; for
selected cases they evaluated both the $\alpha\omega$ and
$\boldsymbol{V}\boldsymbol{\cdot}\boldsymbol{\nabla}\Phi$ formulations and
found their global values identical within error margins. Since the
dry-mass-conserving IFS continuity formulation was introduced only in Cycle
46R1 \cite{Malardel2019}, their Cycle 32R3 calculation used the earlier
source-free total-mass formulation. We infer that the reported agreement is consistent with the model's
source-free total-mass continuity constraint, provided that the pressure-work
variables and mass measure used in the comparison were internally consistent
with that continuity equation. It therefore does not establish physical
gas-phase equivalence of the two formulations. ERA5 was
produced with IFS Cycle 41R2 \cite{Hersbach2020} and likewise predates the
dry-mass-conserving formulation.

JRA-3Q has a closely related structure. It is based on a TL479 version of
JMA's global numerical weather prediction system as of December 2018
\cite{Kosaka2024}. In the JMA Global Spectral Model, the specific-humidity
and cloud-water equations contain tendencies from parameterized processes,
whereas the pressure-mass continuity equation has a zero right-hand side.
Pressure velocity $\omega$ is obtained by integrating this source-free
continuity equation \cite{JMA2019}. Thus model physics can change atmospheric
water variables without the corresponding gas-phase source appearing in the
continuity equation that determines $\omega$. Accordingly, a model-output
category such as ``large-scale'' or ``resolved'' precipitation is not
equivalent to a resolved $\dot{\rho}$ in
Eq.~\eqref{eq:cont_pressure_source}.

MERRA-2 uses a different mass-budget convention. GEOS-5 was modified so that
sources and sinks of atmospheric water enter the atmospheric mass budget while
globally integrated dry atmospheric mass is conserved \cite{Takacs2016}.
This avoids the compensating dry-air variations implied when total atmospheric
mass is held fixed despite changes in water mass. Nevertheless, this is a
continuity budget for total model atmospheric mass, not the gas-phase continuity
equation of Eq.~\eqref{eq:cont_height_source}. In particular, phase changes
between water vapor and suspended cloud condensate do not change total
atmospheric mass, whereas conversion of suspended condensate to precipitation
and re-evaporation of precipitation do enter the MERRA-2 mass source
\cite{Takacs2016}. The three reanalyses therefore need not encode the physical
gaseous source $\dot{\rho}$ in the same way, even before any further processing
of their archived fields.

The distinction between physical and model continuity is separate from another
well-known problem: archived analysis fields need not satisfy the intended
continuity constraint exactly. Such inconsistencies can arise from data
assimilation, interpolation from model to pressure levels, temporal sampling,
lower-boundary treatment, and other processing. \citeA{Trenberth1991} found
large continuity residuals in pressure-level ECMWF analyses and showed that
postprocessing of the analyzed fields can substantially disrupt mass
consistency. Subsequent work developed corrections of the analyzed velocity
field designed to restore mass conservation
\cite{TrenberthHurrellSolomon1995,TrenberthFasullo2018}.

The target of these corrections is physically motivated rather than an
arbitrary numerical closure condition. In the usual approximation that
suspended condensate is neglected in the column mass budget, define the column
masses of water vapor and dry air by
\begin{equation}
  M_v
  \equiv\frac{1}{g}\int_0^{p_s}q_v\,\mathrm{d}p,
  \qquad
  M_d
  \equiv\frac{p_s}{g}-M_v,
  \label{eq:column_dry_water_mass}
\end{equation}
and the corresponding horizontal mass fluxes by
\begin{equation}
  \boldsymbol{F}_v
  \equiv\frac{1}{g}\int_0^{p_s}
  q_v\boldsymbol{u}\,\mathrm{d}p,
  \qquad
  \boldsymbol{F}_d
  \equiv\frac{1}{g}\int_0^{p_s}
  (1-q_v)\boldsymbol{u}\,\mathrm{d}p.
  \label{eq:column_dry_water_flux}
\end{equation}
Physical dry-air conservation and the atmospheric water budget require
\begin{equation}
  \frac{\partial M_d}{\partial t}
  +\boldsymbol{\nabla}\boldsymbol{\cdot}\boldsymbol{F}_d=0,
  \qquad
  \frac{\partial M_v}{\partial t}
  +\boldsymbol{\nabla}\boldsymbol{\cdot}\boldsymbol{F}_v=E-P.
  \label{eq:column_dry_water_continuity}
\end{equation}
Adding the two equations gives the corresponding physical column mass budget,
\begin{equation}
  \frac{1}{g}\frac{\partial p_s}{\partial t}
  +\boldsymbol{\nabla}\boldsymbol{\cdot}
  \left(\frac{1}{g}\int_0^{p_s}\boldsymbol{u}\,\mathrm{d}p\right)
  =E-P.
  \label{eq:physical_column_mass_budget}
\end{equation}
Thus dry-air mass is conserved, whereas total atmospheric mass changes with the
net surface exchange of water. The dry-air constraint can be imposed without
prescribing $E$ and $P$ themselves; together with the moisture budget it makes
the inferred mass balance consistent with, and testable against, independent
estimates of $E-P$.

This distinction is especially relevant when the native model and the
postprocessed analysis obey different constraints. A field can be internally
consistent with a model continuity equation that conserves total mass and yet
fail the physical dry-air constraint in
Eq.~\eqref{eq:column_dry_water_continuity}. Conversely, even a model with a
physically consistent treatment of dry and water mass can acquire additional
continuity residuals through assimilation and archival processing. A diagnosed
continuity residual therefore need not have a single origin.

For uncorrected analysis fields, let
\begin{equation}
  R_d
  \equiv
  \frac{\partial M_d^{\rm a}}{\partial t}
  +\boldsymbol{\nabla}\boldsymbol{\cdot}\boldsymbol{F}_d^{\rm a},
  \label{eq:dry_air_residual}
\end{equation}
where the superscript ${\rm a}$ denotes the original analysis. A common
correction is to add a height-independent horizontal velocity
$\delta\boldsymbol{u}$ so that the corrected dry-air flux satisfies
Eq.~\eqref{eq:column_dry_water_continuity}. For such a barotropic correction,
\begin{equation}
  \boldsymbol{\nabla}\boldsymbol{\cdot}
  \left(M_d^{\rm a}\delta\boldsymbol{u}\right)
  =-R_d.
  \label{eq:barotropic_mass_correction}
\end{equation}
Equation~\eqref{eq:barotropic_mass_correction} represents a simple
column-barotropic correction; more complete three-dimensional adjustment
procedures have also been developed \cite{TrenberthHurrellSolomon1995}. 
The column continuity constraint fixes the divergence of the required mass-flux
correction, but it does not uniquely specify a three-dimensional correction to
the wind field. The choice of a barotropic correction, a vertically varying
correction, or additional constraints on the correction therefore introduces
information beyond mass conservation itself
\cite{Trenberth1991,TrenberthHurrellSolomon1995}.

Pressure velocity can subsequently be reconstructed by vertically integrating
the local continuity equation satisfied by the corrected fields. In general,
if the imposed pressure-coordinate continuity has source $S^{\rm c}$,
\begin{equation}
  \omega^{\rm c}(p)
  =\omega^{\rm c}(p_t)
  +\int_{p_t}^{p}
  \left[
  S^{\rm c}
  -\boldsymbol{\nabla}_{p'}\boldsymbol{\cdot}
  \boldsymbol{u}^{\rm c}
  \right]\mathrm{d}p'.
  \label{eq:omega_from_corrected_continuity}
\end{equation}
The column correction makes the upper and lower boundary conditions compatible,
but the resulting $\omega^{\rm c}$ depends on both the continuity equation being
imposed and the procedure used to correct the horizontal mass fluxes.

The role of such a correction is explicit in the calculation of
\citeA{Laliberte2015}. Their remapping of CESM fields and data assimilation in
MERRA left small continuity errors that could become significant in
energy-flux calculations. They therefore applied a barotropic correction to
the horizontal winds and recovered $\omega$ by integrating the corrected
continuity equation, imposing vanishing $\omega$ at both the upper and lower
boundaries. Their $\alpha\omega$ diagnostic is consequently a quantity conditioned by the
imposed mass-continuity procedure rather than a diagnostic based directly on
an independently archived pressure-velocity field.

There are therefore two conceptually distinct ways in which reanalysis fields
can depart from the physical gaseous continuity equation used in
Section~\ref{sec:theory}. First, the underlying model can employ a mass variable
and source convention that differ from physical gaseous-air continuity, as in
the source-free total-mass formulations discussed above. Second, assimilation,
interpolation, temporal sampling, lower-boundary treatment, and other processing
can introduce additional continuity residuals into the archived fields. A
subsequent mass correction may restore an intended physical conservation
constraint, but it does not uniquely reconstruct the unknown three-dimensional
error in the original velocity field.

These distinctions are especially important for $W_\omega$. As shown in
Section~\ref{sec:theory}, the alpha--omega diagnostic is strongly cancellation
dominated, so comparatively small structured changes in $\omega$ can produce
substantial changes in its global integral. Corrections to horizontal winds can
also alter $W_K$, but $W_K$ is not reconstructed by vertically integrating a
continuity residual and is not intrinsically subject to the same
ascent--subsidence cancellation. Consequently, the reanalysis difference
$W_\omega-W_K$ can reflect the mass-continuity convention of the underlying
model, residual inconsistencies in archived fields, and any subsequent mass
correction, in addition to the physical terms appearing in
Eq.~\eqref{eq:nonhydro_main}. Depending on how continuity and pressure velocity
are treated, the resulting $W_\omega$, and especially its difference from
$W_K$, can therefore vary substantially among reanalyses. This behavior is
examined directly for ERA5, JRA-3Q, and MERRA-2 in the following section.

\section{Data and methods}
\label{sec:methods}
We use MERRA-2 \cite{Gelaro2017}, ERA5 \cite{Hersbach2020}, and JRA-3Q
\cite{Kosaka2024}. Their different models, assimilation systems, grids, and
vertical discretizations provide a test of inter-reanalysis robustness. We
analyze the full calendar year 2025 at 1460 common six-hourly times, applying
identical processing to all three reanalyses for a structural intercomparison
of the two power diagnostics.

We use pressure-level zonal and meridional wind, geopotential or geopotential
height, pressure vertical velocity, temperature, and specific humidity,
together with surface pressure. The files analyzed here contain 37 pressure
levels from 1000 to 1~hPa for ERA5, 42 levels from 1000 to 0.1~hPa for
MERRA-2, and 45 levels from 1000 to 0.01~hPa for JRA-3Q. Product-specific
fill values are converted to missing values before a common physical mask is
applied.

Because the three products extend to different upper pressure levels, we also
tested a common upper boundary at 1~hPa. Truncation at 1~hPa changes $W_K$ by
0.008~W~m$^{-2}$ for MERRA-2 and 0.031~W~m$^{-2}$ for JRA-3Q, and changes
$W_\omega$ by 0.002 and 0.010~W~m$^{-2}$, respectively. These differences do
not affect the conclusions.

Moist specific volume is calculated from virtual temperature,
\begin{equation}
  \alpha_v=\frac{R_dT(1+0.608q)}{p},
  \qquad R_d=287.05\ \mathrm{J\ kg^{-1}\ K^{-1}},
  \label{eq:alpha_virtual}
\end{equation}
where $T$ is temperature, $q$ is specific humidity, $p$ is pressure in pascals,
and $R_d$ is the gas constant of dry air. The local alpha--omega power per unit
mass is
\begin{equation}
  c_{\alpha\omega}=-\alpha_v\omega.
  \label{eq:local_ao_calc}
\end{equation}

On each reanalysis latitude--longitude grid, kinetic-energy generation by
horizontal pressure gradients per unit mass is
\begin{equation}
  k
  =-\frac{u}{a\cos\phi}\frac{\partial\Phi}{\partial\lambda}
  -\frac{v}{a}\frac{\partial\Phi}{\partial\phi},
  \label{eq:k_local_grid}
\end{equation}
where $u$ and $v$ are zonal and meridional wind, $a$ is Earth's radius,
$\lambda$ is longitude, $\phi$ is latitude, and $\Phi$ is geopotential.
Longitude derivatives use a periodic centered difference. For ERA5 and
MERRA-2, centered latitude differences are used and the geographic pole rows
are excluded. JRA-3Q is supplied on nonuniform Gaussian latitudes; its
meridional derivative uses a second-order three-point formula on the actual
latitude coordinates, with one-sided three-point formulas at the two
polewardmost Gaussian rows. A value of $k$ is retained only when the
central point and all four derivative-stencil points are finite and above the
local surface. Geopotential height, where supplied, is converted using
$\Phi=gz$.

For $W_K$, this five-point requirement can raise the effective lowest usable
pressure level above that determined by the local surface pressure alone. A
central point can itself lie above the surface while one or more neighboring
points required for the derivative lie below terrain. This additional masking
does not arise for $c_{\alpha\omega}$, whose evaluation requires no horizontal
derivative stencil. For all three reanalyses, we quantify the stencil effect by
recording, at each pressure level, the area-weighted fraction for which the
central point is valid, the fraction for which the full five-point stencil is
valid, their difference, and the fraction of center-valid area retained by the
stencil. The additional global area excluded by the stencil is confined mainly
to the lowest pressure levels: its maximum annual-mean value is 2.1\% in ERA5,
2.7\% in JRA-3Q, and 3.6\% in MERRA-2, in all cases at 975~hPa. At 1000~hPa,
the stencil retains 97.0\%, 95.9\%, and 94.4\% of the center-valid area,
respectively. The additional loss falls below 0.5\% of global area by 750~hPa
and is negligible by 500~hPa. These quantities are averaged over the 1460
six-hourly analysis times.

For either $x=c_{\alpha\omega}$ or $x=k$, complete archived layers are
integrated as
\begin{equation}
  C_x^{\rm res}
  =\sum_{i=1}^{N-1}
  \frac{x_i+x_{i+1}}{2}\frac{p_i-p_{i+1}}{g},
  \label{eq:resolved_column}
\end{equation}
where $N$ is the number of ordered pressure levels and $i$ is the level index.
A layer is used only when both bounding values are finite and both pressures
satisfy $p_i\leq p_s$.

For $W_K$, the pressure-level and stencil masks leave an unresolved interval
between the lowest valid level $p_1$ and surface pressure $p_s$. We estimate
the sensitivity to this interval in two ways. First, as a zero-surface
sensitivity construction, we set $k_s=0$ and connect this value linearly to
the lowest valid value $k_1$, giving
\begin{equation}
  C_{K,0}^{\rm surf}
  =\frac{k_1}{2}\frac{p_s-p_1}{g}.
  \label{eq:surface_zero_power}
\end{equation}
This construction provides a simple zero-surface treatment against which the
linear extrapolation below can be compared.

As a second estimate, we linearly extrapolate $k$ from the two lowest valid
levels, $p_1>p_2$,
\begin{equation}
  k_s^{\rm lin}
  =k_1+\frac{k_1-k_2}{p_1-p_2}(p_s-p_1),
  \label{eq:surface_extrap_density}
\end{equation}
and calculate
\begin{equation}
  C_{K,\rm lin}^{\rm surf}
  =\frac{k_1+k_s^{\rm lin}}{2}\frac{p_s-p_1}{g}.
  \label{eq:surface_extrap_power}
\end{equation}
Where a second valid lower level is unavailable, the zero-surface estimate
\eqref{eq:surface_zero_power} is used instead. The two treatments characterize
the sensitivity of $W_K$ to the unresolved near-surface layer without assuming
that either reproduces its detailed vertical structure.

We do not include an analogous surface-layer correction in the reported
$W_\omega$. Unlike $k$, the alpha--omega power density has no natural zero
boundary value. At the lower boundary, pressure velocity satisfies
\begin{equation}
  \omega_s
  =\frac{D_s p_s}{Dt}
  =\frac{\partial p_s}{\partial t}
   +\boldsymbol{u}_s\boldsymbol{\cdot}\boldsymbol{\nabla}p_s,
  \label{eq:surface_omega}
\end{equation}
and is therefore generally nonzero at the surface. Consequently,
$c_{\alpha\omega,s}=-\alpha_{v,s}\omega_s$ has no corresponding natural
zero-surface prescription. Nor is there a physical reason to assume that the
strongly sign-varying $c_{\alpha\omega}$ field varies linearly from its lowest
archived pressure level to the surface. We therefore report $W_\omega$ for
complete archived pressure layers only, avoiding an additional lower-boundary
assumption in this already cancellation-dominated diagnostic. Alternative
surface constructions, where calculated, are retained only as numerical
sensitivity tests and do not define the reported $W_\omega$.

Annual global means are obtained by summing the vertically integrated column
contributions with grid-cell-area weighting and dividing by the total surface
area $\mathcal S$; the normalization is therefore not changed by terrain or
stencil masking. Masked contributions are treated as unresolved rather than by
renormalizing the remaining valid area. For JRA-3Q, grid-cell areas are
evaluated using the supplied Gaussian quadrature weights. Zonal means are
formed analogously within each latitude band. Vertical profiles are annual
global means of the archived layer contributions. Robustness is assessed from
synchronized six-hourly correlations, the partition above and below 500~hPa,
and signed latitude profiles.

OpenAI ChatGPT was used as an interactive assistant for manuscript editing,
checking mathematical derivations and analysis code, and reference
verification.

\section{Results}
\label{sec:results}
\subsection{Global integrals}

Table~\ref{tab:headline} summarizes the 2025 annual means. On complete archived
layers, $W_K^{\rm res}=2.36$, 2.48, and 2.84~W~m$^{-2}$ in ERA5, JRA-3Q, and
MERRA-2, respectively. With the zero-surface treatment $k_s=0$, the omitted
surface layer adds 0.23, 0.24, and 0.27~W~m$^{-2}$, giving totals of 2.59, 2.72,
and 3.12~W~m$^{-2}$ and a cross-reanalysis mean of 2.81~W~m$^{-2}$. Linear
extrapolation gives larger surface contributions of 0.51, 0.54, and
0.63~W~m$^{-2}$, corresponding to totals of 2.87, 3.02, and
3.47~W~m$^{-2}$ and a cross-reanalysis mean of 3.12~W~m$^{-2}$.

For comparison, Table~\ref{tab:headline} also gives the 1980--2023 mean
resolved $W_K$ values calculated from the conversion components reported by
\citeA{Oh2026}. They agree closely with the 2025 archived-layer values obtained
here for all three reanalyses.

The alpha--omega diagnostic varies more strongly. Its archived-layer means are
3.23, 1.87, and 2.94~W~m$^{-2}$, while the corresponding values after linear
surface extrapolation are 2.53, 1.28, and 3.20~W~m$^{-2}$. The
cross-reanalysis range therefore increases from 1.36 to
1.92~W~m$^{-2}$. Pairwise correlations of the synchronized six-hourly $W_K$
series are 0.92--0.96 with $k_s=0$ and 0.93--0.96 with linear extrapolation,
compared with 0.69--0.77 for linearly extrapolated $W_\omega$.

\begin{table}
\caption{Atmospheric-power diagnostics for 2025 from 1460 six-hourly analysis
  times. ``Archived'' denotes complete pressure layers. For $W_K$, $C_s^0$ and
  $\mathrm{sum}_0$ use the zero-surface condition $k_s=0$, while
  $C_s^{\rm lin}$ and $\mathrm{sum}_{\rm lin}$ use linear extrapolation to
  $p_s$. For $W_\omega$, only the linear surface treatment is shown. The column
  labelled ``Oh et al.'' gives the 1980--2023 mean resolved $W_K$ obtained as
  $C(P_E,K_E)+C(P_M,K_M)$ by \protect\citeA{Oh2026}; these values do not include a
  surface correction. Values are W~m$^{-2}$.}
\label{tab:headline}
\centering
\small
\setlength{\tabcolsep}{4pt}
\begin{tabular}{lrrrrrrrrr}
\toprule
 & \multicolumn{6}{c}{$W_K$} & \multicolumn{3}{c}{$W_\omega$}\\
Reanalysis & archived & Oh et al. & $C_s^0$ & $\mathrm{sum}_0$
           & $C_s^{\rm lin}$ & $\mathrm{sum}_{\rm lin}$
           & archived & $C_s^{\rm lin}$ & $\mathrm{sum}_{\rm lin}$\\
\midrule
MERRA-2 & 2.84 & 2.87 & 0.27 & 3.12 & 0.63 & 3.47 & 2.94 &  0.26 & 3.20\\
ERA5    & 2.36 & 2.45 & 0.23 & 2.59 & 0.51 & 2.87 & 3.23 & -0.70 & 2.53\\
JRA-3Q  & 2.48 & 2.46 & 0.24 & 2.72 & 0.54 & 3.02 & 1.87 & -0.59 & 1.28\\
\bottomrule
\end{tabular}
\end{table}

\subsection{Latitudinal structure and cancellation}

\begin{figure}[t]
\centering
\includegraphics[width=\textwidth]{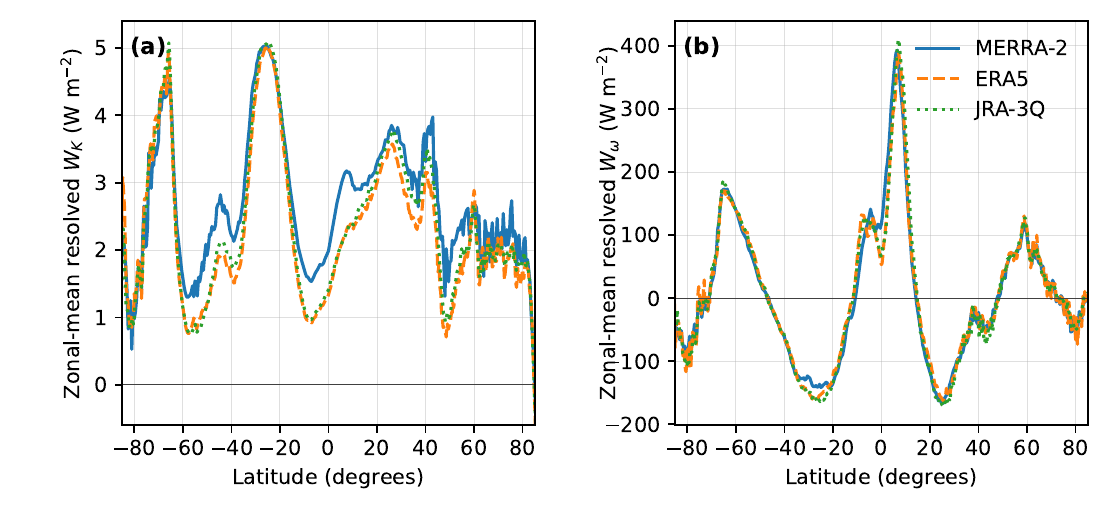}
\caption{Annual zonal-mean archived-layer column power for 2025.
(a) Kinetic-energy generation by horizontal pressure gradients, $W_K$.
(b) Alpha--omega power $W_\omega$. Profiles are shown for
$|\phi|\le85^{\circ}$; the omitted polar bands make net signed contributions
smaller than 0.01~W~m$^{-2}$ to $W_K$ and 0.04~W~m$^{-2}$ to $W_\omega$.
Note the different ordinate scales.}
\label{fig:latitude_profiles}
\end{figure}

The annual zonal-mean profiles reveal the contrasting numerical conditioning
(Fig.~\ref{fig:latitude_profiles}). Kinetic-energy generation by horizontal
pressure gradients is positive through almost the entire latitude range and
closely agrees among the three products. After area weighting, its positive
latitude-band contributions are 2.37--2.85~W~m$^{-2}$, while negative
contributions have magnitudes below 0.01~W~m$^{-2}$. Thus global $W_K$ is not a
small residual of large opposite-signed zonal terms.

The alpha--omega profiles have similar large-scale shapes but much larger
amplitudes. Column values reach about 390--410~W~m$^{-2}$ near $7^{\circ}$N and
$-163$ to $-172$~W~m$^{-2}$ near $25$--$26^{\circ}$N, compared with global
means of only 1.87--3.23~W~m$^{-2}$. Area-weighted positive contributions are
50.4--52.4~W~m$^{-2}$ and negative contributions are
$-47.2$ to $-50.5$~W~m$^{-2}$. The sum of the absolute positive and negative branch contributions therefore
exceeds the global residual by factors of about 30--55.

\subsection{Vertical structure and surface sensitivity}

\begin{figure}[t]
\centering
\includegraphics[width=\textwidth]{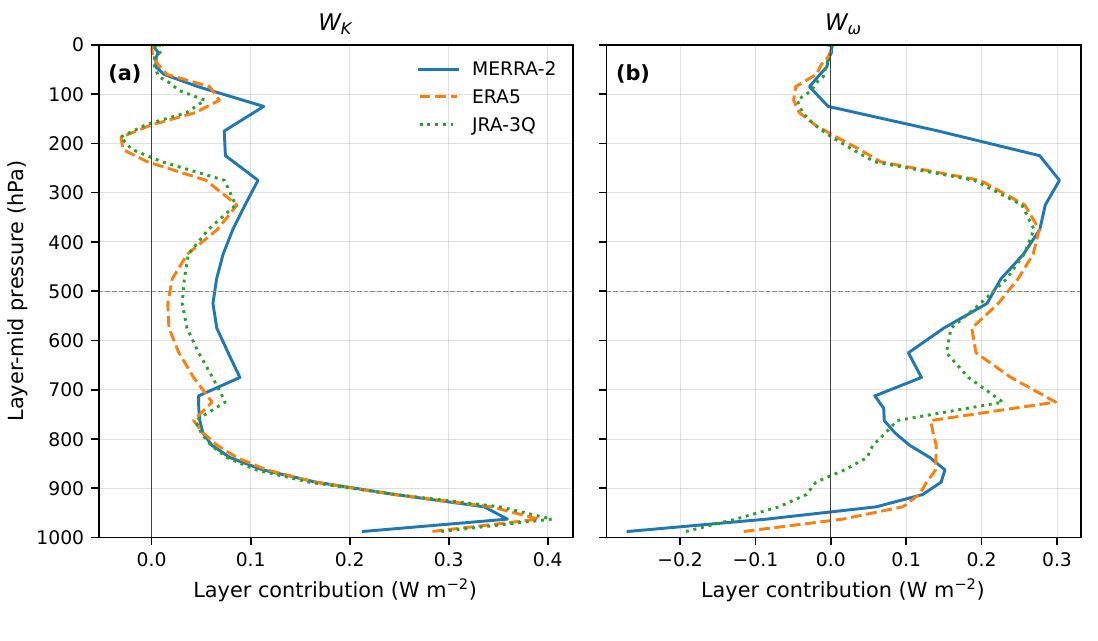}
\caption{Annual global-mean archived-layer contributions for 2025. (a) $W_K$
and (b) $W_\omega$. Pressure decreases upward; the dashed line marks 500~hPa.
Each point is the contribution of one pressure layer. Surface corrections are
not included in the plotted profiles and are reported separately in
Table~\ref{tab:headline}.}
\label{fig:vertical_profiles}
\end{figure}

Horizontal kinetic-energy generation is concentrated in the lower atmosphere
(Fig.~\ref{fig:vertical_profiles}). The largest archived layer occurs at
962.5~hPa in all three products. Below 500~hPa, the archived atmosphere supplies
82.7\%, 82.1\%, and 72.5\% of $W_K^{\rm res}$ in ERA5, JRA-3Q, and MERRA-2. The
corresponding values are 1.95, 2.04, and 2.06~W~m$^{-2}$, a range of only
0.11~W~m$^{-2}$.

The zero-surface treatment adds 0.23, 0.24, and 0.27~W~m$^{-2}$ to $W_K$ in
ERA5, JRA-3Q, and MERRA-2, respectively. These corrections are 9.6--9.7\% of
the archived-layer values and about 8.8\% of the corrected totals. Linear
extrapolation adds 0.51, 0.54, and 0.63~W~m$^{-2}$, raising the archived-layer
values by 21.5--22.1\% and accounting for 17.7--18.1\% of the extrapolated
totals. The difference between the two treatments is 0.28--0.35~W~m$^{-2}$
across the three products, while their similar relative contributions indicate
that the sensitivity is systematic rather than reanalysis specific.

Alpha--omega power is less coherent in the lower atmosphere. Its contribution
below 500~hPa is 2.06~W~m$^{-2}$ in ERA5, 0.76~W~m$^{-2}$ in JRA-3Q, and
1.23~W~m$^{-2}$ in MERRA-2. The contrast is even clearer in the surface
treatment. Linear extrapolation consistently increases $W_K$ in all three
reanalyses, by 0.51, 0.54, and 0.63~W~m$^{-2}$ in ERA5, JRA-3Q, and MERRA-2,
respectively. Applied in exactly the same way to $W_\omega$, however, the
surface correction is $-0.70$, $-0.59$, and $+0.26$~W~m$^{-2}$, respectively:
it decreases $W_\omega$ in ERA5 and JRA-3Q but increases it in MERRA-2.
Thus the near-surface correction has a consistent sign and magnitude for
$W_K$, but is sign-indefinite for $W_\omega$.

On complete archived layers, $W_\omega^{\rm res}-W_K^{\rm res}$ is $+0.87$,
$-0.61$, and $+0.10$~W~m$^{-2}$ in ERA5, JRA-3Q, and MERRA-2, respectively.
For the closure comparison, applying the common linear surface extrapolation to
both diagnostics changes these differences to $-0.34$, $-1.74$, and
$-0.27$~W~m$^{-2}$. The sign-indefinite alpha--omega surface contribution
therefore adds substantial uncertainty to any closure calculation; the
archived and extrapolated differences are interpreted separately below.

\section{Discussion and conclusions}
\label{sec:discussion}
The archived-layer differences $W_\omega-W_K$ are
0.87~W~m$^{-2}$ for ERA5, 0.10~W~m$^{-2}$ for MERRA-2, and
$-0.61$~W~m$^{-2}$ for JRA-3Q. If these quantities were evaluated from a
single internally consistent representation of physical gaseous-air continuity,
Eq.~\eqref{eq:nonhydro_main} would relate their difference to the physical
$W_P$, $W_{\rm nh}$, and boundary and storage terms. Section~\ref{sec:unresolved}
shows why this interpretation cannot in general be applied to reanalysis
pressure-level fields.

The three reanalyses do not encode atmospheric mass continuity in the same way.
ERA5 inherits the source-free total-mass continuity formulation of the
pre-Cycle-46R1 IFS, while the JMA dynamical system underlying JRA-3Q likewise
diagnoses pressure velocity from a source-free pressure-mass continuity
equation. MERRA-2 instead uses a mass-budget formulation in which changes of
atmospheric water affect total atmospheric mass while dry-air mass is conserved.
Moreover, archived pressure-level fields can contain additional continuity
inconsistencies introduced by assimilation, interpolation, temporal sampling,
and lower-boundary treatment. The three values of $W_\omega-W_K$ above should
therefore not be interpreted as three independent estimates of the physical
$W_P+W_{\rm nh}$.

This also changes the interpretation of agreement between the two diagnostics.
The near equality of $W_\omega$ and $W_K$ in MERRA-2 does not imply that the
physical $W_P$ is small. Conversely, the substantial positive ERA5 difference
and negative JRA-3Q difference do not imply corresponding physical source or
nonhydrostatic terms. Likewise, the agreement reported by
\citeA{Steinheimer2008} for an earlier IFS cycle is naturally consistent with
the source-free total-mass continuity equation used by that model generation
and therefore does not establish physical equivalence of the two formulations.

Treatment of the unrepresented surface layer provides an additional numerical
distinction. The alpha--omega surface extrapolation is not robust across the
three reanalyses and makes $W_\omega-W_K$ negative in all three products. We
therefore do not use the surface-extrapolated alpha--omega values to infer a
physical closure. For $W_K$, we instead carry two surface treatments. Imposing
the zero-surface condition $k_s=0$ gives corrections of
0.23--0.27~W~m$^{-2}$ and a cross-reanalysis mean total of
2.81~W~m$^{-2}$, while linear extrapolation gives
0.51--0.63~W~m$^{-2}$ and a mean total of 3.12~W~m$^{-2}$. Thus the treatment
of the omitted layer shifts the inferred mean $W_K$ by about
0.31~W~m$^{-2}$ but does not alter its inter-reanalysis coherence or its order
of magnitude. We retain both surface treatments when estimating total
atmospheric power below.

Although the present analysis is restricted to 2025, its resolved $W_K$ values
closely reproduce the 1980--2023 values calculated from the conversion
components reported by \citeA{Oh2026} (Table~\ref{tab:headline}), with
differences below 4\% for all three reanalyses. 
This supports the use of the 2025 $W_K$ values as representative
of the longer-term magnitude, while the present calculation itself does not
quantify interannual variability.

The cancellation shown in Fig.~\ref{fig:latitude_profiles} helps explain why
$W_\omega$ is especially sensitive to these distinctions. Its global value is
a small residual of much larger positive and negative contributions, whereas
$W_K$ is positive at almost every latitude and agrees much more closely among
the three products. Trenberth-type mass corrections impose a physically
motivated dry-air conservation constraint on analyzed fields, but the column
constraint does not uniquely reconstruct the three-dimensional velocity error.
When $\omega$ is subsequently recovered by vertically integrating the corrected
continuity equation, the resulting field is conditional on both the continuity
constraint and the chosen mass-flux correction. Corrections to horizontal winds
can also alter $W_K$, but $W_K$ is not itself reconstructed by vertical
integration of continuity and is not intrinsically subject to the same
ascent--subsidence cancellation.

Consequently, reanalysis-derived $W_\omega$ is best regarded as a diagnostic of
the particular mass-continuity treatment embodied in, or subsequently imposed
on, the analyzed fields rather than as an independent measurement of physical
atmospheric power. Its difference from $W_K$ can contain the effects of model
mass-continuity conventions, continuity residuals in archived fields, and
postprocessing corrections in addition to the physical terms in
Eq.~\eqref{eq:nonhydro_main}. The large spread of $W_\omega-W_K$ across the
three reanalyses is consistent with this sensitivity.

An observational scale for total $W_P$ can be obtained from the tropical
precipitation-profile estimate. \citeA{PauluisDias2012}, with the correction of
\citeA{PauluisDias2013}, estimated precipitation-induced dissipation at about
1.5~W~m$^{-2}$ for the tropical belt $30^\circ$S--$30^\circ$N from TRMM
precipitation profiles for 1999--2007.
Since this belt occupies one half of Earth's surface, the tropical contribution
alone corresponds to 0.75~W~m$^{-2}$ in the global mean. Using GPCP v2.3 for
2000--2009, \citeA{Nutzel2024} report mean precipitation rates of
3.05~mm~day$^{-1}$ in the same tropical belt and 2.70~mm~day$^{-1}$ globally,
implying an extratropical mean of 2.35~mm~day$^{-1}$. If extratropical
precipitation were assigned the same mean lifting height $H_P$ as tropical
precipitation, the corresponding global $W_P$ would be about
1.33~W~m$^{-2}$. Neglecting the extratropical contribution altogether gives
0.75~W~m$^{-2}$. These deliberately crude estimates place global $W_P$ at
order 1~W~m$^{-2}$. We therefore use $W_P\simeq1$~W~m$^{-2}$ only as an
illustrative scale and emphasize the need for a dedicated global
precipitation-path estimate.

To our knowledge, global $W_{\rm nh}$ is not presently constrained by
observations. If $W_{\rm nh}\geq0$, the two surface-corrected estimates of
resolved $W_K$, together with the illustrative physical
$W_P\simeq1$~W~m$^{-2}$, give cross-reanalysis total-power scales of about
3.8~W~m$^{-2}$ for $k_s=0$ and 4.1~W~m$^{-2}$ for linear extrapolation.
Importantly, this inference uses $W_K$ and the independently estimated physical
$W_P$; it does not use either reanalysis $W_\omega$ or the residual
$W_\omega-W_K$.

This power scale sharpens the contrast with estimates of generation within
the Lorenz framework. \citeA{RomanskiRossow2013}, using largely observation-based temperature and
diabatic-heating fields, reported annual global means
$G_z=1.52$~W~m$^{-2}$ and $G_e=-0.29$~W~m$^{-2}$. Their sum gives
$G_{\rm LEC}=1.23$~W~m$^{-2}$. To our
knowledge, this remains the only global estimate in which APE generation was
directly constrained by observations rather than inferred from reanalysis or
model energetics. The revised radiative-flux-profile product of
\citeA{ZhangRossow2023} now makes such an updated calculation possible, but
it has not yet been reported. The Romanski--Rossow estimate therefore
remains a relevant observational benchmark.

A similarly low power scale was obtained by \citeA{Kleidon2021}, whose
two-box maximum-power model gives about 1.6~W~m$^{-2}$ for kinetic-energy
generation associated with poleward heat transport between tropical and
extratropical regions. This value should not, however, be interpreted as a
general upper bound on kinetic-energy generation by differential heating.
Kleidon's model explicitly excludes kinetic-energy generation by the Hadley
circulation, whereas in the conventional Lorenz energy cycle the Hadley
circulation contributes to the conversion $C(P_M,K_M)$.

The more direct contrast for the present analysis is with $W_K$. Even without
a surface-layer correction, the resolved $W_K$ values of
2.36--2.84~W~m$^{-2}$ substantially exceed the observationally constrained
$G_{\rm LEC}=1.23$~W~m$^{-2}$. This disparity suggests that a substantial
part of kinetic-energy generation diagnosed by $W_K$ is not represented by
conventional Lorenz APE generation. The two surface treatments raise the
cross-reanalysis mean $W_K$ to 2.81--3.12~W~m$^{-2}$; including the
illustrative $W_P\simeq1$~W~m$^{-2}$ gives the corresponding total
atmospheric-power scale of about 3.8--4.1~W~m$^{-2}$.

This discrepancy motivates consideration of additional pathways of
atmospheric power generation not represented by the conventional Lorenz APE
framework. One such possibility has a long history. In 1885, Heinrich Hertz proposed
an atmospheric steam-engine interpretation.
As reported by \citeA{MulliganHertz1997}, Hertz estimated the relevant
evaporation--condensation temperature difference to be about 15~K. For
$T\simeq300$~K, this gives an efficiency scale
$\Delta T/T\simeq0.05$. Applied to the global latent heat flux of order
80~W~m$^{-2}$ corresponding to the GPCP precipitation rates quoted above, this
gives a power scale of about 4~W~m$^{-2}$. The agreement in magnitude with the
atmospheric power inferred here is consistent with a substantial role for water
phase transitions in atmospheric power generation. A more complete
thermodynamic formulation and quantitative assessment of this steam-engine
argument are the subject of ongoing work. Resolving this discrepancy will require
improved observational constraints,
particularly on $W_P$ and $W_{\rm nh}$, together with further theoretical work
on atmospheric power-generation mechanisms.

\section*{Conflict of Interest}

The authors declare there are no conflicts of interest for this manuscript.

\section*{Acknowledgments}
This work received institutional support from the Biotic Pump Greening Group Institute
and the Institute for Advanced Study, Technical University of Munich. No specific grant
funding was received.

\section*{Data and Code Availability}

ERA5 data were obtained from the Copernicus Climate Data Store, using the ERA5 hourly
pressure-level and single-level products \cite{ERA5PressureData,ERA5SingleData}.
MERRA-2 data were obtained from the NASA Goddard Earth Sciences Data and Information
Services Center, collection \texttt{M2I3NPASM} (\texttt{inst3\_3d\_asm\_Np})
\cite{MERRA2M2I3NPASM}. JRA-3Q data were obtained from the NSF NCAR Geoscience Data
Exchange, dataset \texttt{d640000}, using the high-resolution isobaric analysis
(\texttt{anl\_p}) and surface analysis (\texttt{anl\_surf}) products
\cite{JRA3QData}. The derived data and Mathematica scripts used in this study are
archived in Zenodo at \url{https://doi.org/10.5281/zenodo.21966314}
\cite{Zenodo21966314}.

\end{document}